\documentclass[11pt,a4paper]{article}
\usepackage[T1]{fontenc}
\usepackage[latin1]{inputenc}
\usepackage[english]{babel}
\usepackage{amsmath}
\usepackage{amsthm}
\usepackage{amsfonts}
\usepackage{amssymb,amscd}
\usepackage{enumerate}

\usepackage{titling}
\newcommand{\subtitle}[1]{%
  \posttitle{%
    \par\end{center}
    \begin{center}\large#1\end{center}
    \vskip0.5em}%
}
\newtheorem{proposition}{Proposition}

\newtheorem{problem}{Problem}

\newcommand{\set}[1]{\left\{#1\right\}}

\newcommand{\rev}[1]{{#1}^{\text R}}

\newcommand{\A}{\mathcal{A}}

\newcommand{\EXTRA}[1]{}

\newcommand{\defin}[1]{\emph{#1}} 

\usepackage{authblk}

\title{Another proof of undecidability for the correspondence decision problem}
\subtitle{Had I been Emil Post}
\author[1,2]{Vesa Halava\thanks{vesa.halava@utu.fi\\ MS Classification: 03D35, 03D03, 68Q01}}
\affil[1]{\small Department of Mathematics and Statistics, University of Turku, Finland}
\affil[2]{\small Department of Computer Science, University of Liverpool, UK}

\begin{document}

\maketitle


\begin{abstract}
In 1946  Emil Leon Post (\emph{Bulletin of Amer.
Math. Soc.} {\bf 52} (1946), 264 -- 268) defined a famous correspondence decision problem which is nowadays called the Post Correspondence Problem, and he proved that the problem is undecidable. In this article we follow the steps of Post, and give another, simpler and more straightforward  proof of the undecidability of the problem using the same source of reduction as Post original did, namely, the Post Normal Systems.
\end{abstract}

\section{Introduction}

The original formulation of the \emph{Post Correspondence Problem} (or, as Post called it, \emph{correspondence decision problem}),  PCP for short, by Emil
Post~\cite{Po-46} is the following:

\begin{problem}[Post Correspondence Problem]\label{PCP2}
Let $B=\set{a,b}$ be a binary alphabet, and denote by $B^*$ the set of all finite words over $B$. Given a finite set of $n$ pairs
of words,
$$
W=\{(u_i,v_i)\mid u_i,v_i\in B^*, \ i=1,2,\dots,n \}.
$$
Does there exist a nonempty sequence
$i_1, i_2,\dots , i_k$ of indices, where each $i_j\in \set{1,2,\dots, n}$ for $1\le j \le
k$, such that
\begin{equation}\label{eq:preli2}
u_{i_1}u_{i_2}\cdots u_{i_k}=v_{i_1}v_{i_2}\cdots v_{i_k}\,?
\end{equation}
\end{problem}

In the history of computation, the Post Correspondence Problem and its variants have played
a major role as a simply defined algorithmically undecidable problems that can be
used to prove other undecidability results. Here we concentrate on the undecidability proofs of the PCP itself.
In his article \cite{Po-46}, Post proved that the problem is unsolvable, or undecidable, as we say today,  by a technical and nontrivial reduction from the \defin{assertion problem of the Post normal systems}.  We shall  give another proof for the undecidability of the PCP from the same source.

A standard textbook proof of the PCP's undecidability employs the undecidability of the \emph{halting problem of the Turing machines} as the base of reduction, see for example \cite{Si-12}, or the construction by Claus \cite{Cl-80} from the \emph{word problem of the semi-Thue systems} to the PCP that gives the best known undecidability bounds for $n$ in the definition of the PCP. The number $n=|W|$ of the pairs of words in an \emph{instance} $W$  of the PCP is called the \emph{size} of $W$.

The standard reduction from the Turing machines or semi-Thue system to the PCP have a common idea: An instance of the PCP is constructed in  a way that any solution to it is a (possibly coded) concatenation  of all configurations of a required computation or derivation of the original machine or system. This is not the case in Post's original proof of undecidability, indeed,  he uses only the words in the rules of an instance of normal system. A sequence of these rule words imply a required derivation in the Normal system, if and only if the sequence is a solution of the instance of the PCP. The new proof presented in this article is based on the idea of a standard type: a solution exists to the constructed instance of the PCP, if and only if the solution is a concatenation of the full configurations required of the given Post normal system. 

Finally, in Post's definition the PCP is defined for binary words. Actually, the cardinality of the alphabet $B$ is not relevant, since every instance of the PCP with any
alphabet size has an equivalent one in terms of binary words using an injective encoding into binary alphabet $\set{a,b}^*$ from $B^*$. For
example, if $B=\set{a_1,a_2,\dots,a_k}$, then $\varphi$ defined by
$\varphi(a_i)=a^ib$ is such an encoding.

\section{Normal systems}

We give a formal definition of a normal system
instead of the bit informal one used by Post in \cite{Po-46}.

Let
$A=\set{a,b}$ be a binary alphabet, and let $X$ be a variable ranging over words in $A^*$.
A \defin{normal system}
$S=(w,P)$ consists of a \defin{initial word} $w \in A^+$ and a finite set $P$ of
\defin{rules} of the form 
$\alpha X \mapsto X\beta$,
where $\alpha,\beta\in A^*$. We say that a word $v$ is a
\defin{successor} of a word $u$, if there is a rule
$\alpha X\mapsto X\beta$ in $P$ such that $u=\alpha u'$ and
$v=u'\beta$. We denote this by $
u\to v$.
Let $\to^*$ be the reflexive and transitive closure of $\to$. Then
$u\to^* v$ holds if and only if $u=v$ or there is a finite sequence of
words $u=v_1, v_2,\dots, v_n=v$ such that $v_i\to v_{i+1}$ for 
$i=1,2,\dots, n-1$.
The Post normal systems are a special case of the \emph{Post canonical systems} for which Post proved in 1943 the Normal-Form Theorem, see \cite{Po-43}.

The \defin{assertion} of a normal system $S=(w,P)$ is the set
\begin{equation}
\mathcal {A}_S=\set{v\in A^*\mid w \to^* v}\,.
\end{equation}

The following undecidability result is cited in \cite{Po-46} to
Post~\cite{Po-43} and for the formal proof there is a reference to Church~\cite{Ch-43}.

\begin{proposition}\label{prop:ns}
It is undecidable for a given  normal system $S=(w,P)$ and a word
$u\in A^+$, whether or not $u\in \mathcal A_S$.
\end{proposition}

Actually, the problem remains undecidable even if we assume that
in each rule $\alpha X\mapsto X\beta$ in $P$ the words $\alpha$ and
$\beta$ are non-empty, and therefore, this is assumed in the
following. We shall call the problem, asking for a given word $u$, whether or not $u\in \mathcal A_S$ the \defin{assertion problem}.

\section{The proof by Post}

The idea of the Post's original undecidability proof is the following: Assume that $u\in \mathcal A_S$, where $S=(w,P)$ and let
\begin{equation}\label{eq:preli1}
w=\alpha_1x_1,\ x_1\beta_1=\alpha_2x_2,\dots,
x_{k-1}\beta_{k-1}=\alpha_k x_k, \ x_k\beta_k=u\,,
\end{equation}
where $\alpha_j X \to X\beta_j$ and $x_{j}\in A^*$ for all $j$ and $k>0$. Post proves that  existence of a sequence in \eqref{eq:preli1} is equivalent to the following two conditions
\begin{equation}\label{eq:preli3}
w\beta_{1}\beta_{2}\cdots \beta_{k}=\alpha_{1}\alpha_{2}\cdots \alpha_{k}u\, 
\end{equation}
and
\begin{equation}\label{eq:preli5}
|w\beta_{1}\beta_{2}\cdots
\beta_{{j-1}}|\ge|\alpha_{1}\alpha_{2}\cdots
\alpha_{j}|\, , \text{for all }j=1, \dots, k,
\end{equation}
where $|v|$ denotes the \emph{length of the word} $v$. In other words, it is proved that  \eqref{eq:preli3} and \eqref{eq:preli5} are
equivalent to the condition $u\in \mathcal A_S$.

The rest of Post's constructions is the transformation of the system $S$ to a form where the equation \eqref{eq:preli5} holds if \eqref{eq:preli3} holds. Post does this by introducing a new symbol $c$, considering the reverse words and adding cyclic shifts of all words in $\mathcal A_S$ to the assertion of the system. Namely, the normal system
 $S_1=(\rev {w}c,P_1)$
where
\[
P_1=\set{\rev{\alpha}cX\mapsto Xc\rev{\beta} \mid
\alpha X\mapsto X\beta \in P}\cup \set{yX\mapsto Xy\mid y\in \set{a,b,c}}
\]
is constructed. Next Post proves that $u\in \mathcal A_S$ if and only if there are rules $\gamma_{j}X\mapsto X\delta_{j}$ in $P_1$ such that
\begin{equation}\label{eq:preli6}
\rev{w}c\delta_{1}\delta_{2}\cdots
\delta_{k}=\gamma_{1}\gamma_{2}\cdots
\gamma_{k}\rev{u}c\,,
\end{equation}
and  that the length condition of the form \eqref{eq:preli5} is true for \eqref{eq:preli6}. Indeed, occurrences of the marker symbol $c$ guarantee that the length condition of the form \eqref{eq:preli5} is satisfied. The reverse words and conjugate rules are added in order to making it possible to work with marked rules.

Finally, Post uses \eqref{eq:preli6} to produce an instance of the PCP. He applies a trick  called \emph{desynchronization}; let $d$ be a new symbol and define two mappings  $\ell_d$ and $r_d$  from $\{a,b,c\}^*$ to $\{a,b,c,d\}^*$ such that, for each word $v=a_1a_2\cdots a_t$ with $a_i\in \{a,b,c\}$, $\ell_d(w)=da_1da_2\cdots da_t$ and $r_d(w)=a_1da_2d\cdots a_td$.
Now $u\in \mathcal A_S$ if and only if
there exists a solution for the instance
\begin{equation}
\set{(\ell_d(\delta),r_d(\gamma))\mid \gamma X \mapsto X \delta \in P_1}\cup
\set{(d\ell_d(\rev{w}c),d),(dd,r_d(\rev{u}c)d)},
\end{equation}
of the PCP.
Indeed, by desynchronization, a solution to the PCP must begin with $(d\ell_d(\rev{w}c),dd)$, and end
with $(dd,r_d(\rev{u}c)d)$.  Post concludes, by
Proposition~\ref{prop:ns}, that the PCP is undecidable.


\section{New proof}

As Post, we start with the sequence \eqref{eq:preli1}, but use different indeces, that is, assume that there exists a sequence
\begin{equation}\label{eq:uud}
w=\alpha_{i_1}x_1,\ x_1\beta_{i_1}=\alpha_{i_2}x_2,\dots,
x_{k-1}\beta_{i_{k-1}}=\alpha_{i_k} x_k, \ x_k\beta_{i_k}=u\,,
\end{equation}
for a normal system $A=(w,P)$ and input word $u$ where $\alpha_{i_j}X\mapsto X\beta_{i_j}\in P$ for $j=1,\dots, k$. Instead of equations \eqref{eq:preli3} and \eqref{eq:preli5}, we take
\begin{equation}\label{eq:mun}
wx_1\beta_{i_1}x_2\beta_{i_2} \cdots x_k\beta_{i_k} = \alpha_{i_1}x_1\alpha_{i_2} x_2 \cdots \alpha_{i_k} x_k u,
\end{equation}
where all configurations of the sequence in \eqref{eq:uud} are concatenated in two ways.
Let $c$ and $f$ be new letters and assume that the cardinality of the production set $P$ is $t$ and  denote $P=\{p_1,\dots,p_t\}$ where $p_j = \alpha_{j}X\mapsto X\beta_{j}$ for $j=1,\dots, t$. For every $p_j\in P$, we define two pairs of words,
$$
p_{j}^{\alpha}=(\ell_d(c^jf), r_d(f\alpha_{j})) \quad \text{ and }\quad p_{j}^{\beta}=(\ell_d(\beta_j), r_d(c^j )).
$$
where $r_d$ and $\ell_d$ are the desynchronizing mappings for a new letter $d$. In other words, we  split all productions of $P$ into two pairs. The word $c^jf$ is a marker word forcing us to chose these pairs jointly in a solution of an instance of the PCP defined next. Now, define an instance of the PCP by the pair of words
\begin{equation}
\begin{split}
W=& \{(d\ell_d(fw),dd),(dd,r_d(fu)d), (da,ad), (db, bd)\} \\
& \cup  \{p_{j}^{\alpha},p_{j}^{\beta} \mid j=1,\dots, t\}.
\end{split}
\end{equation}
It is straightforward to prove that $u\in \A_S$ if and only if there exists a solution to the PCP. Indeed, all the solutions to the instance of the PCP are of the form
\begin{equation}
\begin{split}
&d\ell_d(fwc^{i_1}f x_1\beta_{i_1}c^{i_2}f \cdots c^{i_k}f x_k \beta_{i_k})dd \\
&=d\ell_d(fw) \ell_d(c^{i_1}f) \ell_d (x_1) \ell_d(\beta_{i_1}) \ell_d(c^{i_2}f)  \cdots \ell_d(c^{i_k}f)\ell_d (x_k) \ell_d(\beta_{i_k}) dd \\
&=
dd r_d(f\alpha_{i_1})r_d(x_1) r_d(c^{i_1}) r_d(f\alpha_{i_2}) r_d(x_2)  \cdots r_d(f\alpha_{i_k}) r_d(x_k) r_d(c^{i_k}) r_d(fu) d \\
&=ddr_d(f\alpha_{i_1}x_1c^{i_1}f \alpha_{i_2}x_2c^{i_2}f\cdots \alpha_{i_{k}}x_{k} c^{i_k}f u)d
\end{split}
\end{equation}
implying sequences of the form  \eqref{eq:uud} for the given normal system $S$. 

Finally, note that we are forced to the split the rules in two pairs as the words $x_i$  appear in different sides of the words $\alpha_i$ and $\beta_i$ in \eqref{eq:uud} and, therefore, $\alpha_i$ and $\beta_i$ cannot be set in a common pair of words.

\section{Conclusion}

A new, shorter and bit simpler proof for the undecidability of the PCP was given, using the same source of undecidability, the Post normal systems, as was used in the original proof by Post. Indeed, this new proof could have been found by Post as well, but as a true pioneer of the field of computations he immediately would have noticed the following deficiency of the construction: when considering the size of an instance of the PCP constructed, Post's original construction gives an instance of size  $|P|+5$, but our new construction gives an instance of size $2|P|+4$. As the undecidable problem in the normal system, the cardinality of $P$ must be at least two, we realize that  Post's proof gives a better bound for the undecidability. Therefore, I could not have done anything better - had I been Emil Post.

\paragraph{Acknowledgement.} Author thanks Professor Tero Harju for excellent comments.

\end{document}